\documentclass[referee]{aa}
\usepackage{graphicx}

\begin{document}

\title{The radio counter-jet of the QSO 3C~48}
\author{  W. X. Feng    \inst{1,2}
       \and T. An      \inst{1,3}
       \and X. Y. Hong  \inst{1,3}
       \and Jun-Hui Zhao \inst{4}
       \and T. Venturi \inst{5}
       \and
       \\ Z. -Q. Shen  \inst{1}
       \and W. H. Wang  \inst{1,3}}%
\institute{Shanghai Astronomical Observatory, Chinese Academy of
Sciences, Shanghai 200030, China \and Graduate School of  Chinese
Academy of Sciences, Beijing 100039, China \and National
Astronomical Observatories, Chinese Academy of Sciences, Beijing
100012, China \and Harvard-Smithsonian Center for Astrophysics, 60
Garden Street, MS 78, Cambridge, MA 02138 \and Istituto di
Radioastronomia del CNR, via Gobetti 101, 4129 Bologna, Italy}

\offprints{W. X. Feng, wfeng@shao.ac.cn}
\date{Received $<$date$>$ / Accepted $<$date$>$}

\abstract {We present multi--frequency radio observational results
of the quasar 3C~48. The observations were carried out with the
Very Large Array (VLA) at five frequencies of 0.33, 1.5, 4.8, 8.4,
and 22.5 GHz, and with the Multi--Element Radio Linked
Interferometer Network (MERLIN) at the two frequencies of 1.6 and
5 GHz. The source shows a one--sided jet to the north within
1\arcsec,  which then extends to the northeast and becomes
diffuse. Two bright components (N2 and N3), containing most of the
flux density are present in the northern jet. The spectral index
of the two components is $\alpha_{N2}\sim-0.99\pm0.12$ and
$\alpha_{N3}\sim-0.84\pm0.23$ ($S\propto\nu^{\alpha}$). Our images
show the presence of an extended structure surrounding component
N2, suggestive of strong interaction between the jet and the
interstellar medium (ISM) of the host galaxy. A steep--spectrum
component, labelled as S, located 0.25$''$ southwest to the
flat--spectrum component which could be the core of 3C 48, is
detected at a significance of $>15\sigma$. Both the location and
the steepness of the spectrum of component S suggest the presence
of a counter--jet in 3C 48.

{}}

\maketitle\markboth{The radio counter-jet of the QSO 3C~48} {W. X.
Feng, et al.}

\maketitle

\section{Introduction}

The radio source 3C 48 (B0134+329) is a 16th magnitude quasar,
with redsift $z=0.368$ (Spinrad et al. \cite{Spinrad85}). It was
the first QSO identified at optical wavelengths (Smith et al.
\cite{Smith61}), and the second identified one (after 3C 273) at
radio wavelengths (Schmidt \cite{Schmidt63}). The source shows
unexpectedly strong emission in the far IR bands, suggesting that
it is located in a gas--rich environment (Chatzichristou et al.
1999). The variability measured with ROSAT suggests that at least
part of the X--ray emission (principally the softest) arises from
the region of the accretion--disc  or from a beamed blazar
component (Worrall et al. \cite{Worrall04}). The image obtained
with University of Hawaii's 2.2 meter optical telescope at Mauna
Kea Observatory is dominated by a previously unsuspected high
surface brightness region, which was referred to as ``3C 48A'',
centered $\sim1\arcsec$ ($1\arcsec$ $\sim 4.1$ kpc of projected
linear size \footnote{Hubble constant H$_0$=75km s$^{-1}$
Mpc$^{-1}$, deceleration parameter q$_0$=0.5 and zero cosmological
constant}) to the north of the QSO nucleus ( Stockton \& Ridgway
\cite{Stockton91}; Hook et al. \cite{Hook94}). This region was
identified as the potential second nucleus of a galaxy in the
process of merging with the host galaxy of 3C 48 (Zuther et al.
\cite{Zuther03}). Evidence indicates that the nuclear activity in
3C 48 is triggered by the merger process.

3C~48 is classified as a Compact Steep Spectrum (CSS) source based
on its steep radio spectrum and its small angular size (Fanti et
al. \cite{Fanti85}). The spectrum is reasonably straight from
$\sim178$~MHz up to $2.7$~GHz, with a integrated radio index
$\alpha\sim-0.83$ (Peacock et al. \cite{Peacock82}). It shows a
low--frequency peak at about 100~MHz
(\texttt{http://nedwww.ipac.caltech.edu}). The quasar 3C~48 does
not show significant variations with time and have a simple
spectrum at MHz/GHz wavelengths (Baars et al. \cite{Baars77}), and
it is usually used as the primary flux density calibrator in
interferometer observations owing to its compactness and
brightness. While Ott et al. 1994 found that the source could have
become brighter from 1960s to 1990s, the increase in flux density
is enhanced at shorter wavelengths, as is consistent with the
activity in the core (Wilkinson et al. \cite{Wilkinson90}).

\begin{table*}
\centering \caption{Observational information and image
parameters} \vspace{2mm}
\begin{tabular}{cccccc|ccll}\hline\hline

No &Epoch  &Array & Freq.&BW$^a$&$\tau^b$&$S_{peak}$&\emph{r.m.s.}&Maj$\times$Min, PA  &Figure \\
   &       &      & (GHz)&(MHz) &(minute)&(Jy/b)    &(mJy/b) &(arcsec, $\degr$)        &       \\
(1)& (2)   & (3)  & (4)  & (5)  &  (6)   &(7)       &(8)     & (9)                     &(10)   \\\hline
1  &1992.45&MERLIN& 5.0  & 13   &  850   &  2.61    &   0.3  &$0.04\times0.04$, 34.4   &Fig.1d\\\hline
2  &1993.64&MERLIN& 1.6  & 15   &  900   &  4.65    &   1.3  &$0.17\times0.14$, 33.2   &Fig.1b\\\hline
3  &1999.08&VLA-C & 8.4  & 50   &    2   &  3.17    &   0.6  &$2.37\times2.26$,  7.9   &      \\
4  &       &      & 22.5 & 50   &    7   &  1.02    &   0.5  &$1.07\times1.03$, $ 10.6$&      \\\hline
5  &2000.92&VLA-A & 8.4  & 50   &    2   &  1.66    &   0.3  &$0.32\times0.20$, $-73$  &      \\
6  &       &      & 22.5 & 50   &  5.5   &  0.34    &   0.4  &$0.12\times0.07$, $-72$  &Fig.1c\\\hline
7  &2003.46&VLA-A & 0.33 &0.003 &  4.5   &  43.0    &   15   &$11.3\times6.2$,  $-77.0$&      \\
8  &       &      & 1.5  &0.012 &  4     &  15.3    &   1.8  &$2.1\times1.3$,   $-81.8$&      \\
9  &       &      & 4.8  & 50   &  3     &  3.69    &   1.0  &$0.46\times0.33$, $-63.3$&      \\
10 &       &      & 8.4  & 50   &  4     &  1.60    &   0.3  &$0.27\times0.21$, $-82.5$&Fig.1a\\\hline \hline

\end{tabular}\label{tab:obs}\vspace{2mm}\\
\raggedright $^a$ the VLA observations at 0.33 and 1.5 GHz were
carried out in spectral line mode, and
recorded with narrow bandwidth;\\
$^b$ on-source integration time.
\end{table*}

Most of the radio emission occurs on a scale smaller than about
$0.6\arcsec$ and additional emission extends to $\sim6\arcsec$
northwest to the nucleus, with a clumpy morphology (Hartas at al.
\cite{Hartas83}; Woan \cite{Woan92}). At sub--arcsecond
resolution, the source exhibits an elongated structure in the
north--south direction. VLBI observations showed a disrupted radio
morphology (Wilkinson et al. 1991). The active nucleus, which is
identified from its inverted radio spectrum (Simon et al.
\cite{Simon90}), is weak at radio frequencies and is buried deeply
within the host galaxy. A one--sided jet stretches out of the core
towards the north $\sim1\arcsec$ within the nucleus of the host
galaxy. The northern jet is highly dispersed at
0.05\arcsec--0.5\arcsec, and its diffused morphology
 has been interpreted as the result of the jet outflow
interaction with the dense ISM in the host galaxy (Wilkinson et
al. \cite{wilkinson91}). There is also morphological evidence for
a recent merger event, such as a possible double nucleus and a
tidal tail extending several arcseconds to the northwest
(Chatzichristou et al. \cite{Chat99}; Canalizo et al.
\cite{Canalizo00}; Zuther et al. \cite{Zuther03}).

In this paper, we present results of the multi--frequency VLA and
MERLIN observations of 3C~48. The observations and data reduction
are  described in the next section. The analysis of the images is
given in Section 3. Section 4 is a discussion. A summary is given
in the last section.

\section{Observation and data reduction}

Quasar 3C~48 was observed as a flux density calibrator with the
VLA on Jan. 29, 1999 (observational code is AH635) at 8.4 and 22.5
GHz, on Dec. 1, 2000 (AH721) at 8.4 and 22.5 GHz, as well as on
June 17, 2003 (AZ143) at 0.33, 1.5, 4.8 and 8.4 GHz. The
observations were carried out in snapshot mode. 3C 48 was observed
at 4.8, 8.4 and 22.5 GHz with the VLA in the C configuration with
a bandwidth of 50~MHz in two IFs,  while the observations at  0.33
and 1.5 GHz were done in the spectral line mode in order to reject
radio frequency interference (RFI). Two epochs of archival MERLIN
data at 1.6 and 5.0~GHz were also used in this paper. The
observational information of the VLA and the MERLIN is listed in
Table 1 in a time series. Column 1 is the serial number. Columns 2
and 3 are the epoch and the array configuration. The observing
frequency, bandwidth, and on--source time are listed in columns 4,
5, and 6, respectively.

The raw VLA data were calibrated and corrected for the phase and
amplitude errors using the Astronomical Image Processing System
(AIPS) of the National Radio Astronomy Observatory (NRAO). The
MERLIN data were calibrated with the suite of D-programs
(\cite{THO}).

The further process in the reduction of the VLA and MERLIN data,
including editing, self-calibration and image analysis, was
performed using both the AIPS and the DIFMAP packages (Shepherd,
Pearson \& Taylor 1994). In order to do analysis of the jet
structure, we fitted Gaussian models to the self-calibrated
\emph{uv} data--sets. The task MODELFIT in the DIFMAP package was
used for all  data--sets except for the 5~GHz MERLIN, where we
used the task JMFIT in AIPS in order to take out the extended
emission prior to the model fitting. We found that 80--90\% of the
total flux density of the source is dominated by a few compact
components. Elliptical Gaussian models were preferentially used.
If elliptical Gaussian models failed in the least--square fitting,
then a circular Gaussian was used. The errors in the model fitting
parameters are estimated based on the method described by Fomalont
(\cite{Foma99}). The uncertainties of flux density, originating
from the calibration precess, are estimated within a range between
5\% (for VLA) and 10\% (for MERLIN). The uncertainty of the
component position is proportional to the FWHM of the synthesized
beam and is inversely proportional to the ratio of peak signal to
noise (Fomalont \cite{Foma99}).

\begin{figure*}
\resizebox{\hsize}{!}{\includegraphics{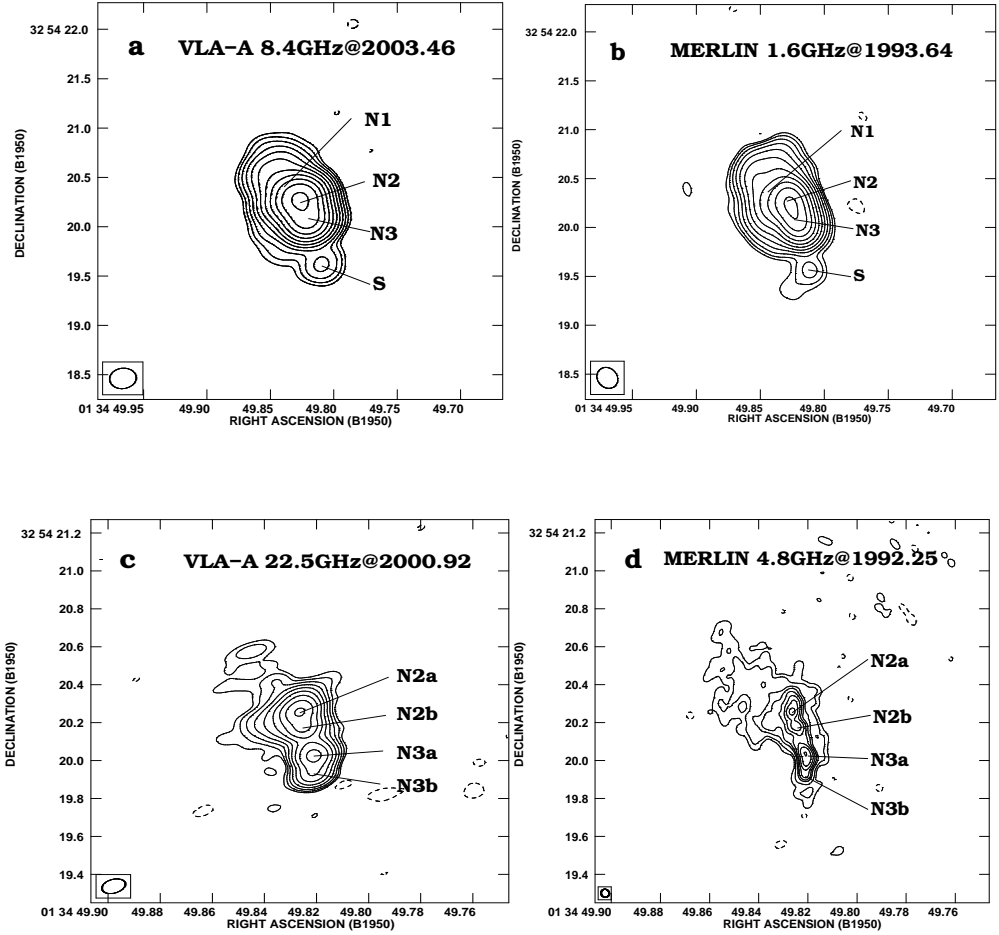}} \caption{The
multi--frequency total intensity images of 3C 48. The peak flux
density, beam size, off source \emph{r.m.s} level are referred to
in columns 7--9 in Table 1. Contours levels for Figs {1a, 1b, and
1c } are 1.1, 5.0, and 1.2 mJy/b$\times$($-1$, 1, 2,..., 1024),
respectively, and those for Fig. 1d  are 1.1 mJy/b$\times $($-1,$
1, 4, 16, 32, 64, 128, 256, 512). The lowest contour levels are
generally 3 times of the \emph{r.m.s} levels.} \label{figure1}
\end{figure*}

\section{Analysis}
\subsection{Images}
We imaged the source using the data--sets listed in Table 1.  The
results show that the source is resolved  when the resolution is
higher than 0.5\arcsec. Fig. 1 shows four images at 1.6, 5, 8.4
and 22.5 GHz with sub--arcsec resolutions. The detailed structure
of the source is shown in the images from Fig.1a to 1d based on an
order of increasing resolution. These images are made using
uniform weighting. Since most of the observations lack nearby
phase calibrators, the bright, compact component N2 was used as
the phase reference in the self--calibration and imaging process.
The parameters of each image are summarized in Table 1. Columns 7
and 8 give the peak flux density and the off--source \emph{r.m.s.}
noise in each of the images; column 9 gives the size of the
synthesized beam; column 10 indicates the label of the images that
are present in Fig. 1. Each of the major components in the images
are fitted in least square to a Gaussian model. The results of
model--fit to the calibrated data are listed in Table 2. Column
(1) is the serial number of the observations, which is the same as
that in Table 1; column (2) is the observing epoch; column (3) is
the observing frequency; column (4) is the component label; column
(5) is the integral flux density of Gaussian component; columns
(6) and (7) are the peak position of the Gaussian component in
RA$_{B1950}$ and Dec$_{B1950}$ coordinates; columns (8), (9), and
(10) are the sizes of  major and minor axes and the position
angle. Previous observations show that the source is dominated by
steep spectrum components and the radiation from the core is
rather weak  and is difficult to identify at radio frequencies. We
note that the phase center in each image is closer to the
brightest jet component N2 (RA$_{B1950}$: $01^h34^m49.826^s$,
DEC$_{B1950}$: $+32\degr54\arcmin20.260\arcsec$) than the radio
core. The core component and its location will be discussed in
Section 4.1.

\begin{table*}
\centering \caption{Parameters of  gaussian components}
\vspace{2mm}
\begin{tabular}{ccccrrcccc}\hline\hline

No& Freq.  &Comp.&$S_{int}$&RA$\;\;\;\;\;\;\;\;\;\;\;\;$&DEC$\;\;\;\;\;\;\;\;\;\;$             &a     & b   &$\Phi$ &Figure  \\
  &  (GHz) &     &(Jy)     &($^h,\;^m,\;^s$)$\;\;\;\;\;\;\;\;\;$&($\degr,\;\arcmin,\;\arcsec$)$\;\;\;\;\;\;\;\;\;$ &(mas)&(mas)&(\degr)&\\
(1)& (2)   &(3)  &(4)      &(5)$\;\;\;\;\;\;\;\;\;\;\;\;$&(6)$\;\;\;\;\;\;\;\;\;\;\;\;$&(7)&(8)&(9)&    \\\hline\hline

1 &4.8     & N2a & 1.549  &01 34 49.826$\pm$0.005  & 32 54 20.264$\pm$0.004   &86.0  &53.6 &168.1  &Fig.1d  \\
  &        & N2b & 0.799  &      49.826$\pm$0.004  &       20.175$\pm$0.003   &65.0  &57.3 &71.8   &        \\
  &        & N3a & 1.423  &      49.821$\pm$0.006  &       20.025$\pm$0.003   &99.7  &47.4 &11.0   &        \\
  &        & N3b & 0.164  &      49.821$\pm$0.003  &       19.936$\pm$0.002   &57.6  &43.5 &1.1    &        \\\hline

2 &1.6     & N1  & 3.29   &      49 840$\pm$0.034  &       20.296$\pm$0.036   &512.0 &342.5 &4.9   & Fig.1b  \\
  &        & N2  & 10.33  &      49.827$\pm$0.030  &       20.264$\pm$0.017   &255.5 &224.71&179.8 &        \\
  &        & N3  &  3.13  &      49.821$\pm$0.006  &       20.075$\pm$0.006   &264.0 &205.3 &      &        \\
  &        & S   & 0.039  &      49.812$\pm$0.026  &       19.574$\pm$0.014   &187.2 &99.4  &172.0 &        \\\hline
3 &8.4     & N   & 3.20   &      49.826$\pm$0.036  &       20.264$\pm$0.010   &317.2 &105.4 &25.4  &        \\\hline
4 &22.5    & N   &1.06    &      49.827$\pm$0.055  &       20.260$\pm$0.021   &262.0 &84.2  &18.9   &\\\hline

5 &8.4     & N1  &  0.211 &      49.843$\pm$0.030  &       20.512$\pm$0.030   &330.0 &330.0&       &       \\
  &        & N2  &  1.997 &      49.827$\pm$0.012  &       20.281$\pm$0.012   &131.3 &131.3&       &        \\
  &        & N3  &  1.025 &      49.821$\pm$0.011  &       20.075$\pm$0.011   &118.3 &118.3&       &        \\
  &        & S   &  0.013 &      49.807$\pm$0.033  &       19.619$\pm$0.026   &356.0 &280.0&112.9  &        \\\hline

6 &22.5    & N2a &  0.55  &      49.827$\pm$0.038  &       20.263$\pm$0.038   &64.1  &64.1 &       &Fig.1c  \\
  &        & N2b &  0.35  &      49.825$\pm$0.045  &       20.172$\pm$0.045   &75.7  &75.7 &       &        \\
  &        & N3a &  0.30  &      49.821$\pm$0.033  &       20.029$\pm$0.033   &56.5  &56.5 &       &        \\
  &        & N3b &  0.12  &      49.821$\pm$0.028  &       19.940$\pm$0.028   &48.4  &48.4 &       &        \\\hline

7 &0.33    & N   &43.14   &      49.827$\pm$0.081  &       20.257$\pm$0.058   &466.4 &332.7&-40.9  &      \\\hline
8 &1.5     & N   &15.88   &      49.826$\pm$0.022  &       20.260$\pm$0.006   &373.5 &108.0&28.4   &    \\\hline
9 &4.8     & N   &5.35    &      49.826$\pm$0.014  &       20.254$\pm$0.020   &322.7 &146.1&24.6   &      \\
  &        & S   &0.015   &      49.814$\pm$0.018  &       19 420$\pm$0.016   &705.5 &341.1&80.6   &      \\\hline
10&8.4     & N1  &0.250   &      49.839$\pm$0.031  &       20.453$\pm$0.031   &330.0 &330.0&       &Fig.1a  \\
  &        & N2  &1.848   &      49.827$\pm$0.011  &       20.280$\pm$0.011   &118.3 &118.3&       &        \\
  &        & N3  &1.011   &      49.821$\pm$0.010  &       20.079$\pm$0.010   &107.5 &107.5&       &        \\
  &        & S   &0.016   &      49.810$\pm$0.027  &       19.633$\pm$0.027   &292.0 &282.0&72.7   &        \\\hline\hline
\end{tabular}\\
\end{table*}

Fig. 1a is the VLA image at 8.4~GHz observed at the epoch of 2003.
The image shows an elongated structure in the NE--SW direction,
which includes 4 components (N1, N2, N3, and S). The main jet is
dominated by the components N1, N2, and N3. Component S has been
detected for the first time at a significance level above
$15\sigma$, at RA $\sim01^h34^m49.810^s$ and DEC
$\sim32\degr54\arcmin19.626\arcsec$. An independent VLA image at
8.4 GHz,  made from the observation at the epoch 2000, confirms
the detection. The epoch 2000 image at 8.4 GHz is not shown, while
the parameters derived from model fitting are listed in Table 2.

The two major components (N2 and N3), the northern extended
component (N1), and the weak southern component (S) are also
detected in the 1.6~GHz MERLIN image (Fig.1b). A Gaussian fit
failed in fitting the diffuse component N1, visible in the MERLIN
1.6 GHz image. Component S is confirmed at 12$\sigma$, and can be
fitted with a Gaussian of
 $0.28\arcsec\times0.23\arcsec$ (PA $=37.6\degr$) with
peak brightness of $26$ mJy per beam.

Both components N1 and S appear to be completely resolved in the
high resolution images at 22.5 GHz (Fig. 1c) observed with VLA and
at 5 GHz observed with MERLIN (Fig. 1d). At the high resolutions,
the component N2 and N3 are further resolved into subcomponents,
labelled as N2a and N2b, N3a and N3b, respectively.

\begin{figure*}
\includegraphics[width=0.5\hsize,angle=0]{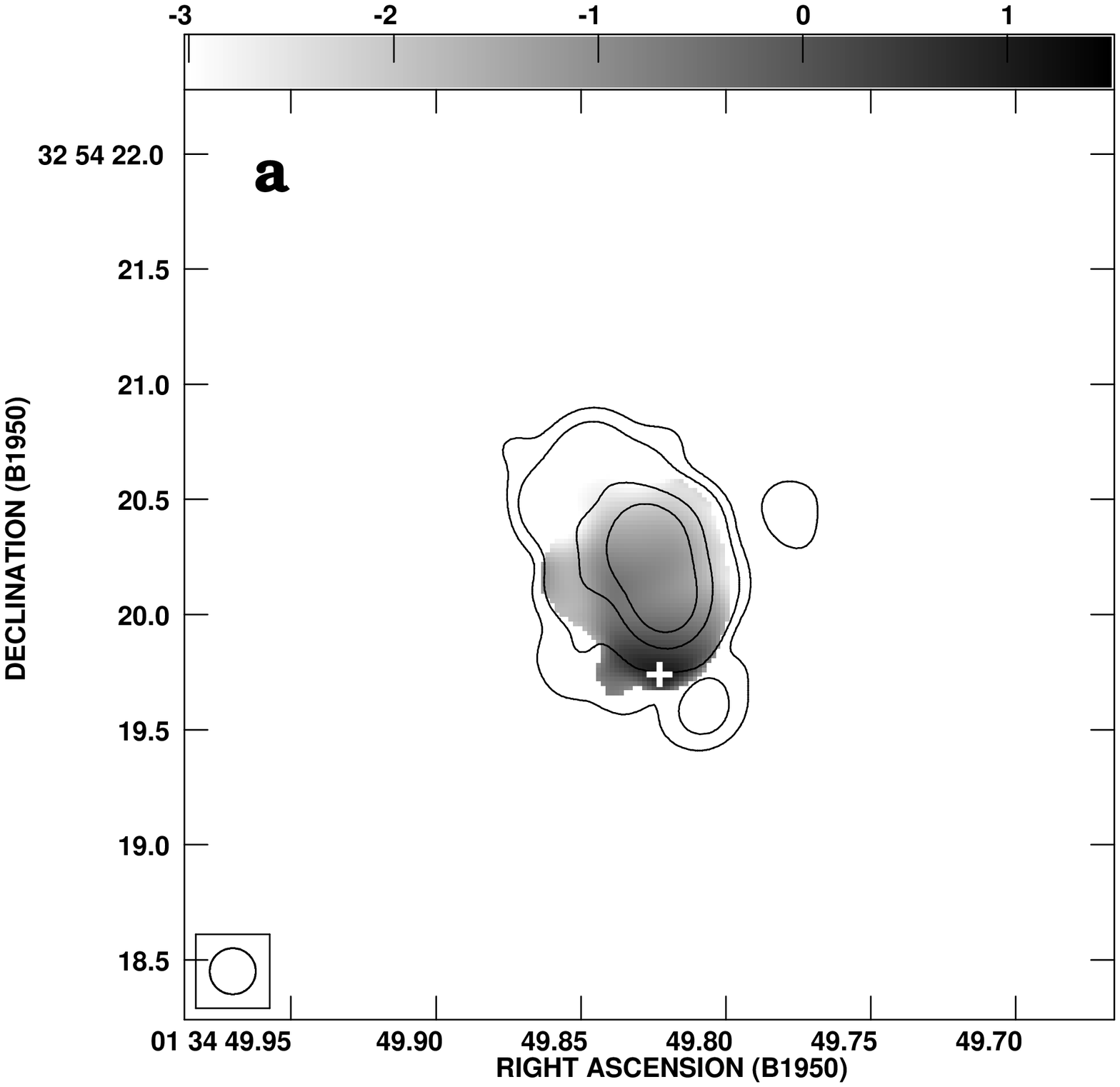}
\includegraphics[width=0.5\hsize,angle=0]{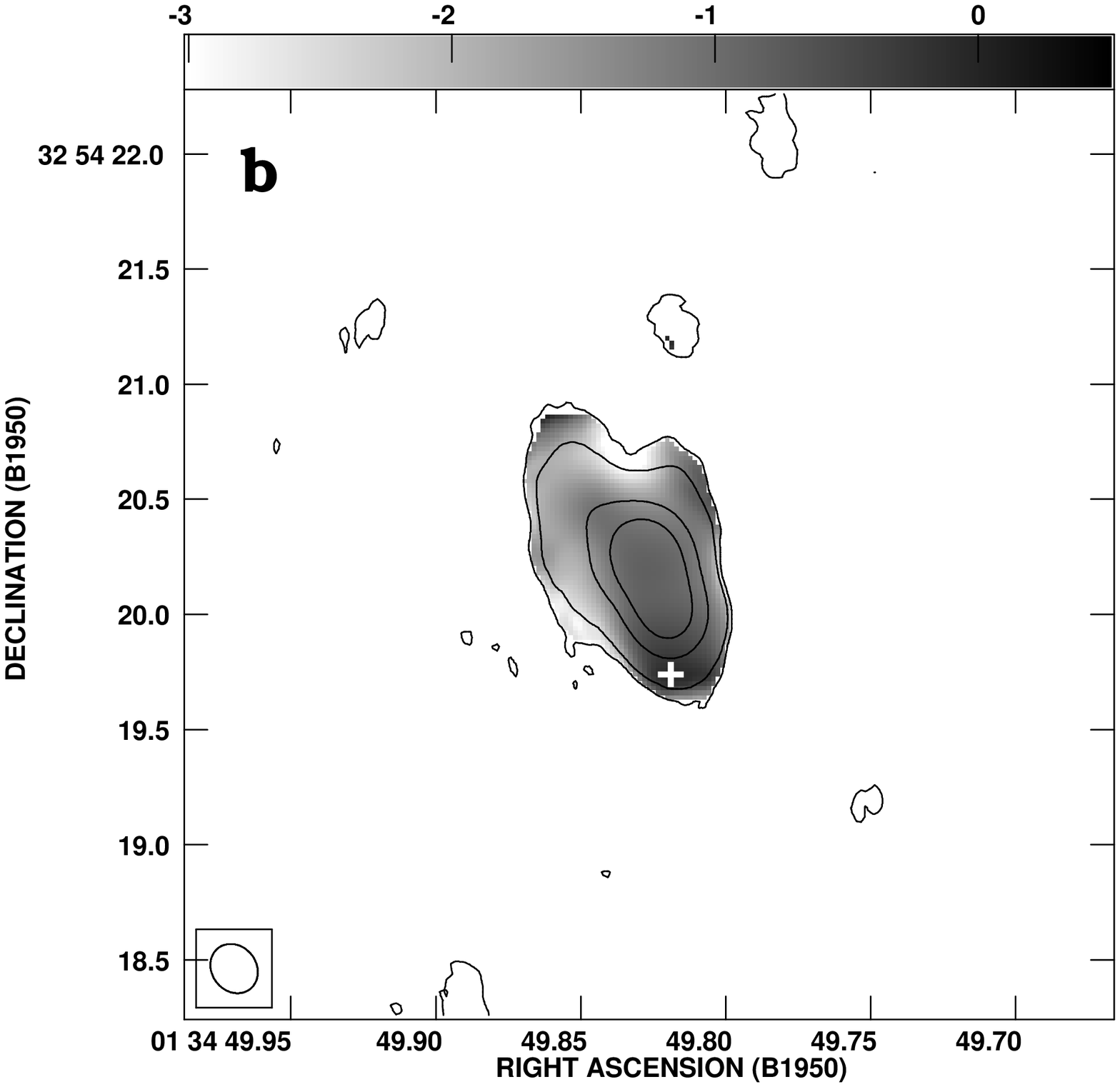}
\caption{{\bf a)} Simultaneous spectral index distribution (grey)
between 8.4 and 22.5 GHz at the epoch 2000.92, superimposed on the
22.5 GHz VLA image. {\bf b)} Spectral index distribution (grey)
between 1.6 and 5 GHz, superimposed on the 5 GHz MERLIN image at
the epoch 1992.45; Restored beam of the contour images are
$0.20\arcsec\times0.20\arcsec$ and $0.23\arcsec\times0.19\arcsec,$
p.a.$ $ $45\degr$; The contour levels are $1.3 mJy/b\times (-1, 1,
4, 64, 256)$ and $3.5 mJy/b\times(-1, 1, 4, 64, 256)$,
respectively. The scale of the spectral index is shown in each
panel. The darkest region corresponds to the flattest spectrum.}
\label{figure2}
\end{figure*}

\subsection{Spectral index}

The multi--frequency observations allow us to study the spectral
index distribution of 3C 48 at sub arc--second resolutions. Fig.
2a shows the image of the  spectral index distribution (grey
scale) derived from the VLA images at 8.4 and 22.5 GHz at the
epoch of 2000.92, with the contours of the radiation intensity at
8.4 GHz superimposed. The uv--data at 22.5 GHz is tapered to match
the synthesized beam of the data at 8.4 GHz. Both images at 8.4
and 22.5 GHz are convolved to the same resolution and are aligned
with respect to the common phase center. With the same method, we
obtained the spectral index image (Fig.2b) with the MERLIN data
sets. Fig. 2b shows the spectral index image (grey scale) between
1.6 and 5 GHz, with the MERLIN 5 GHz contours overplotted. The
dark end of the grey scale corresponds to the flat spectrum. Most
of the source shows a steep spectrum. The spectrum is steepest in
the northern jet (component N2),  while it becomes flatter to the
south.

In order to image the distribution of the spectral index across
the source, we plotted the spectral index along the jet axis at
position angle of $8.5\degr$. The slice spectral indexes are shown
in Figs.3a and 3b, corresponding to Figs.2a and 2b, respectively.
There is a component centered at RA $\sim
01^h34^m49.820^s\pm0.020^s$, DEC
$\sim32\degr54\arcmin19.740\arcsec\pm0.013\arcsec$, marked with a
cross in Figs.2a\&b, showing rather flat spectrum.

The spectral index of components N2, N3 and S are estimated based
on the model fitting results. Fig.4 shows the power--law fitting
in logarithmic coordinate. The spectrum of component N2
($\alpha_{N2}\sim-0.99\pm0.12$) is slightly steeper than that of
component N3 ($\alpha_{N3}\sim-0.84\pm0.23$); and component S
shows also a steep spectrum with $\alpha_{s}\sim-0.70\pm0.14$
between 1.6 and 8.4 GHz. The results are consistent with those
shown in Figs. 2 and 3. A mean spectral index
$\alpha\sim-0.87\pm0.10$ for the overall source (3C 48) is
estimated from the total flux densities at the frequencies between
0.33 and 22.5 GHz (Fig. 4).

\begin{figure*}
\resizebox{0.49\hsize}{!}{\includegraphics{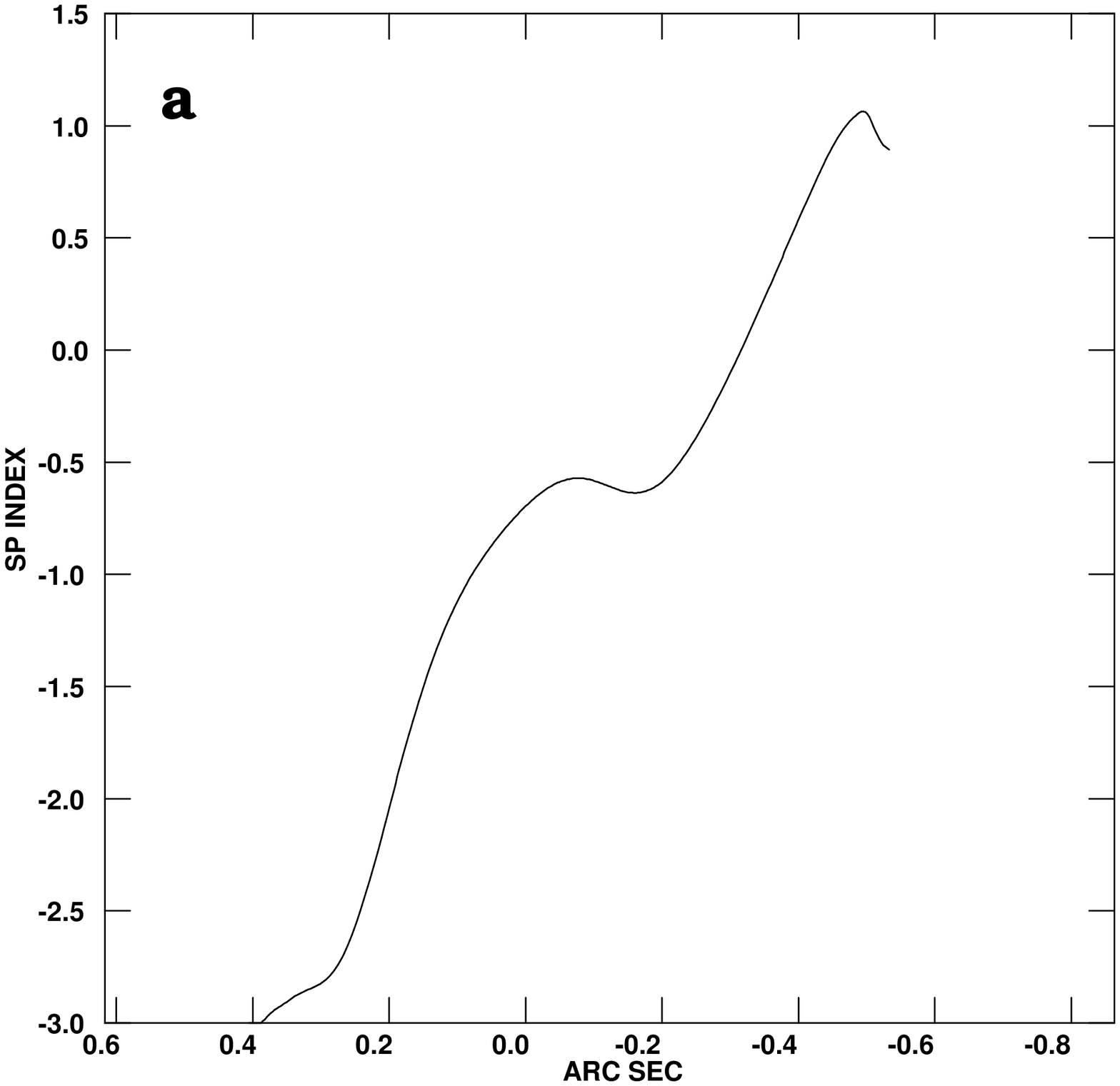}}
\resizebox{0.49\hsize}{!}{\includegraphics{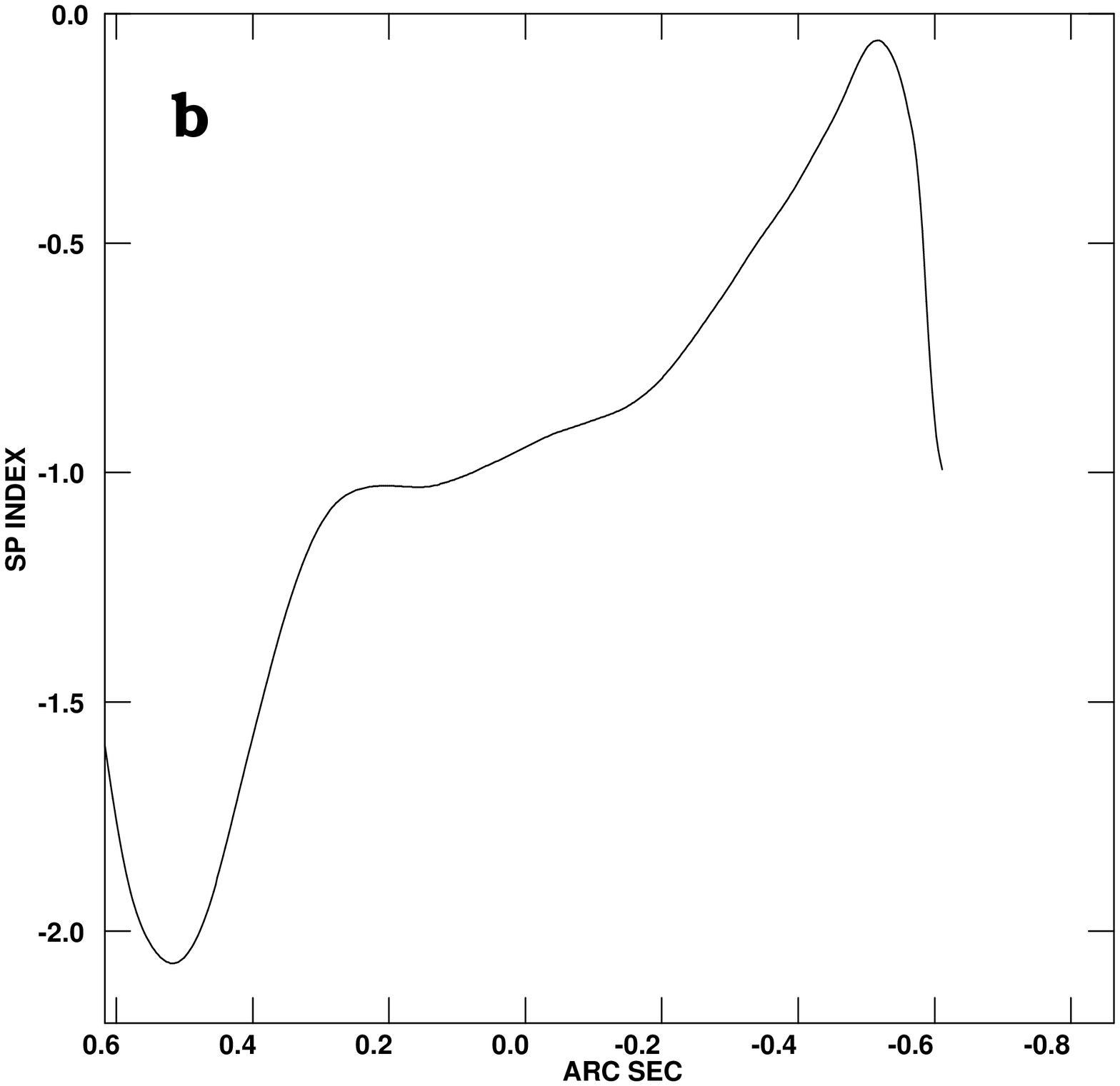}}
\caption{Slice plot of the spectral index at PA 8.5\degr
corresponding to Figs.2a and 2b, respectively. The slice center
for {\bf a)} is RA $01^h34^m49.827^s$, DEC
$+32\degr54\arcmin20.250\arcsec$ and for {\bf b)} RA
$01^h34^m49.827^s$, DEC $+32\degr54\arcmin20.268\arcsec$,
respectively.} \label{figure3}
\end{figure*}

\section{Discussion}
\subsection{The position of the nucleus }

In Section 3.2 we showed that the flat--spectrum component
centered at RA $\sim 01^h34^m49.820^s\pm0.020^s$,  DEC
$\sim32\degr54\arcmin19.740\arcsec\pm0.013\arcsec$, is most likely
the nuclear core in 3C 48. The flat--spectrum core is likely to be
buried in the strong emission at the centimeter and longer
wavelengths. A detection of the active core was also claimed by
Simon et al. (1990). The core was identified at the southern end
of the source in their 22.5 GHz image. Wilkinson et al. (1991)
suggested that the radio core detected by Simon et al. 1990 is
associated with the southernmost component in their high
resolution image. Based on the analysis of our data, we find that
the position of the flat--spectrum core in our image is
$0.15\arcsec$ south of that identified by Simon et al. 1990 and
Wilkinson et al. 1991. Further higher resolution and higher
dynamics range images at millimeter wavelengths are needed to
unambiguously identify the nucleus in 3C 48.

\subsection{A counter--jet in 3C 48}
We have distinctly detected an extended component S
$\sim0.25\arcsec$ southwest of the flat--spectrum core, which is
the putative nucleus of 3C 48. Component S shows a steep spectrum,
i.e. $\alpha_{s}\sim-0.70\pm0.14$. Due to its location and the
steep spectrum nature, we suggest that S is a counter--jet.

Alternatively, a multi--particle model  has been successfully
developed and has been applied to the host galactic nucleus of
3C\,48 (Scharw\"achter et al. \cite{Scharw04}), suggesting a
counter tidal tail extending to the southwest in front of the main
body of the 3C 48 host. The detection of component S could be a
supportive evidence for the counter tidal tail in this quasar.

\begin{figure}
\resizebox{\hsize}{!}{\includegraphics{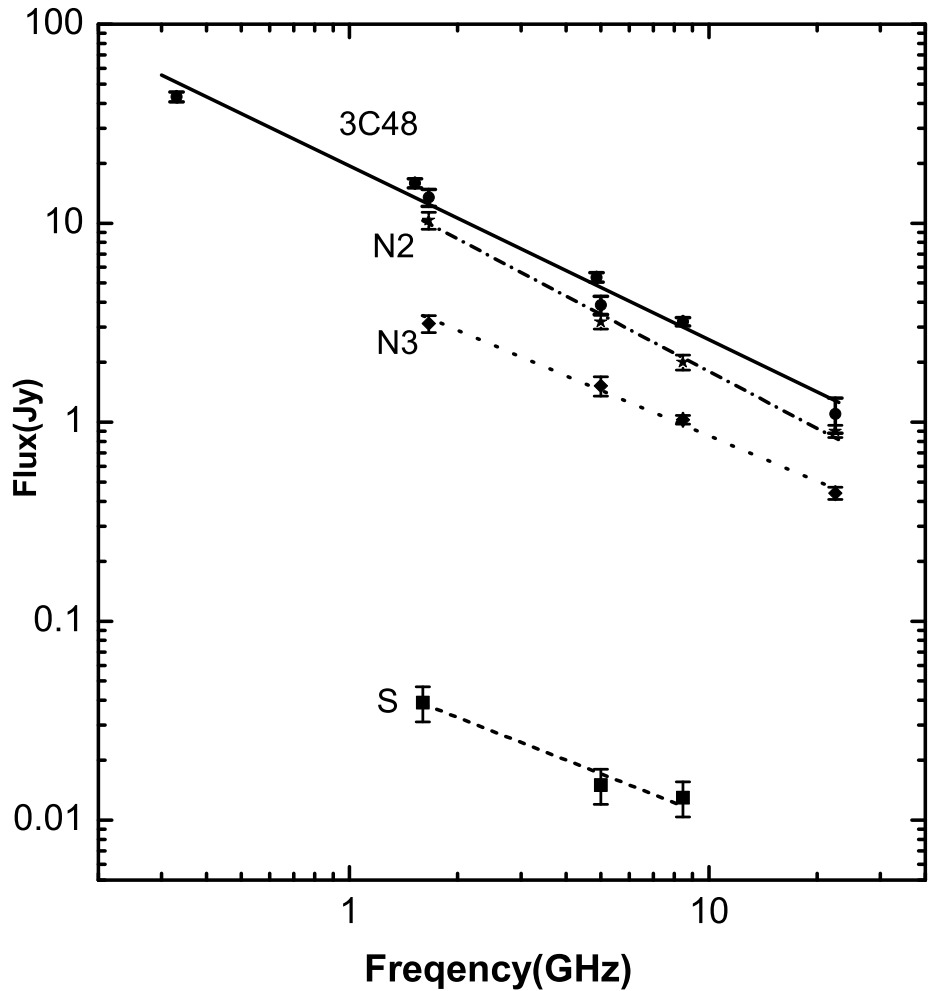}}\\
\caption{Total spectrum of 3C 48 (solid line) and of the various
components in the source.}

\label{figure4}
\end{figure}

\subsection{Interaction with ISM}
CSS sources usually have sub-kpc sizes with complex radio
morphologies. The nature of CSS sources is not completely
understood yet and two scenarios are suggested. In the {\it youth
scenario}, CSS sources are small and could reflect an early stage
in the evolution of radio sources (Phillips \& Mutel
\cite{Phillips82}, Fanti et al. \cite{Fanti95}). According to the
{\it frustration scenario}, , the unusual conditions in the
interstellar medium of their host galaxies, such as a higher
density and/or the presence of turbulence, inhibit the radio
source from growing to larger sizes (van Breugel et al.
\cite{Van84}). The youth scenario is very much preferred with
statistical research (Saikia et al. \cite{Saikia95},
\cite{Saikia01}; Murgia et al. \cite{Murgia99}, \cite{Murgia01}).
Despite this statistical interpretation, the frustration scenario
may still apply to individual sources which can be interpreted in
terms of confinement and interaction between the jets and ISM in
the host galaxies. Numerical simulations have also shown that a
dense ISM may substantially confine CSS sources (De Young
\cite{De93}). The radio structure of 3C~48 has shown the evidence
for strong interaction between the radio jet and the ISM in the
host galaxy. Here is a working scenario. A supersonic collimated
jet stretches out from the core to the north where a shock is
produced in  the interaction between jet flow and the dense
medium. The interaction converts the kinetic energy carried by the
jet to internal energy or turbulence which could enhance the
acceleration of particles and then the radiation field, causing
the jet to brighten. The observed north jet brightening (hot spot
N2 in Fig.1c and Fig.1d) after losing collimation appears to be
consistent with this picture. Based on the study of the kinematics
and physical conditions of the extended emission line gas and
their relations to the nuclear star formation, Chatzichristou
(2001) found that the coupling between radio jet and gas is the
kinematic signature of  a recent interaction. Both Canalizo \&
Stockton (2000) and  Chatzichristou et al. (1999) suggested that
the unusually high stellar velocity dispersion as well as very
young stellar populations are possibly related  to the interaction
of the nuclear radio jets with the ambient media. The NIR (near
infrared) images and spectra also suggest that reddening by
several  magnitudes in the nuclear emission can be explained by
the interactions of the 3C 48 radio jet with the circum--nuclear
medium (Zuther et al. \cite{Zuther03}). The ratio of the
recombination lines $P_{a}\beta$ and $P_{a}\gamma$ determined from
3C 48 (Zuther et al. \cite{Zuther03}) is about 1.6 ($\pm0.5$),
which is consistent with the value typical for the ISM of active
galactic nuclei being excited by the central--engine continuum
emission.

\section{Summary}

In this paper, we carried out radio multi--frequency observations
of 3C 48. The high dynamic range images exhibit the detailed
structure of the source on the sub--arcsec scale. The position of
the nucleus with a flat spectrum is determined based on the
spectral index distribution images. The jet components show steep
spectra. We confirm the existence of a counter--jet in 3C 48,
which shows the evidence for  the first time that 3C 48 actually
has a double-sided jet. The northern jet appears to be disrupted
at 0.5\arcsec northeast to the core. Our analysis suggests that
the initially collimated jet can be disrupted by the interaction
between jet flow and the circum-nuclear medium.

\section{Acknowledgement}

This research is supported by the National Science Foundation of
PR China (10328306, 10333020, and 10473018). The authors are
grateful for the technique support by the staff of the VLA and the
MERLIN. The VLA is a facility of the National Radio Astronomy
Observatory, which is operated by Associated Universities Inc.
under cooperative agreement with the National Science Foundation.
MERLIN is a National Facility operated by the University of
Manchester at Jodrell Bank Observatory on behalf of PPARC.
W.X.Feng thanks the MERLIN staff (Peter Thomasson) for providing
the archive data. It is a great pleasure to thank the referee Paul
Wiita who suggested some improvements,  and also thanks Prof.
D.R.Jiang for his valuable advices in this paper. This research
has made use of the NASA/IPAC Extragalactic Database (NED).

\end{document}